# Approaching Emergent Risks: An Exploratory Study into the Risk Management of Artificial Intelligence Systems in Financial Sector Organisations

Finlay McGee

**Abstract -** Globally, artificial intelligence implementation is growing, holding the capability to fundamentally alter organisational processes and decision making. Simultaneously, this brings a multitude of emergent risks to organisations, exposing vulnerabilities in their extant risk management frameworks. This necessitates a greater understanding of how organisations can position themselves in response. This issue is particularly pertinent within the financial sector with relatively mature AI applications matched with severe societal repercussions of potential risk events. Despite this, academic risk management literature is trailing behind the speed of AI implementation. Adopting a management perspective, this study aims to contribute to the understanding of AI risk management in organisations through an exploratory empirical investigation into these practices. In-depth insights are gained through interviews with nine practitioners from different organisations within the UK financial sector. Through examining areas of organisational convergence and divergence, the findings of this study unearth levels of risk management framework readiness and prevailing approaches to risk management at both a processual and organisational level. Whilst enhancing the developing literature concerning AI risk management within organisations, the study simultaneously offers a practical contribution, providing key areas of guidance for practitioners in the operational development of AI risk management frameworks.

# 1. Introduction

## 1.1. Background

The implementation of Artificial Intelligence (AI) within the contemporary organisational landscape is burgeoning. AI has diffused globally, pervading organisational sizes and industries and holding seismic growth projections in coming years (McKinsey, 2023). The multivariate capabilities of AI afford transformative potential and the possibility of reshaping businesses and society at every level. Yet, as with any novel technology, AI presents an array of novel and emergent risks. Much like the potential of AI, these risks have pervasive implications for organisations and wider society.

Whilst there is no universally agreed definition for AI, it can be broadly defined as 'the theory and development of computer systems that are able to perform tasks that normally require human intelligence' (Galimova et al., 2019). Within this work, AI is used as a blanket term to refer to the extent of techniques through which this occurs, including machine learning (ML), natural language processing (NLP), and computer vision. The generalisable nature of AI technologies provide an extensive applicability within a range of use cases and contexts.

As a historic pioneer of novel technologies, the financial sector (FS) is one of the most prolific adopters of AI (Herrmann and Masawi, 2022). In 2022, the BoE and FCA (2022) found 72% of UK FS organisations in the process of designing or implementing AI systems, with adoption likely to triple in the coming years. The FS encompasses a range of organisations including banks, financial services (investment banks, asset management, and financial advisory etc.) and insurance companies. Within the sector, AI is driving the emergence of novel 'mechanisms, innovations, models, products and services' (Cao, 2022: p.2). Owing to this, AI applications in the FS are expansive, ranging from backend implementation like robotic process automation and consumer onboarding to financial applications such as mathematical modelling and financial advice systems (OECD, 2021).

**1.2 Rationale and Aims**

The risk management (RM) of emerging technologies and information systems are seen as increasingly pertinent issues in the modern digitised society (Bandyopadhyay, Mykytyn and Mykytyn, 1999; Luo, 2022), especially within the context of the FS. This is magnified with the emergence of AI which poses severe, complex and pervasive risks to organisations and society and threatens existing RM approaches (Cheatham, Javanmardian and Samadari, 2019). Despite the wealth of literature framing the risks of organisational AI from a social perspective, research surrounding the risks of AI from an organisational perspective is comparably limited (Wirtz, Weyerer and Kehl, 2022). This leaves scholars and practitioners lamenting the lack of robust governance and RM controls for AI systems and calling for greater academic insight (Canhoto and Clear, 2020; Baquero, 2020; Kurshan, Shen and Chen, 2020; Eitel-Porter, 2020; Hu et al., 2021).

Whilst AI RM frameworks are emerging (NIST, 2023), academic literature is struggling to keep pace with the speed of AI adoption. Robust empirical investigation into RM approaches is needed to inform the advance of practical frameworks. This is especially important in the case of the FS with relatively extensive AI applications and severe consequences of potential risk events (Bartneck et al., 2021). Despite this, empirical work into the topic is limited (Wirtz, Weyerer and Kehl, 2022). This study aims to contribute by conducting in-depth exploratory insight into AI RM practices within FS organisations.

**1.3 Scope and Implications**

Through interviews with nine practitioners, this study examines AI adopting organisations within the UK FS. Through this, it aims to bring timely insight into their approaches to AI risks and AI RM on both a processual and organisational level. This study has clear organisational implications. From this perspective AI risks are creating a hindrance to its adoption and resulting in implementation failures (Westenberger, Schuler and Schlegel, 2021; Zhang et al., 2022). Thus, a greater empirical insight into RM of AI facilitates better RM of AI in practice. However, due to consistencies of AI applications and their resultant RM necessities, the insights can hold foundational relevance in a wider organisational context. From an overarching perspective, better organisational RM of AI extends beyond the organisations themselves, as organisational risks can directly impact upon consumers and society. Thus, improving RM at the micro-level is fundamental to limiting the possibility of macro-level harm.

## 2. Literature Review

The intention of this literature review is to frame the research topic whilst dissecting the contemporary literature around AI RM practices in organisations and the FS. As robust research requires rigid foundational concepts (Grant and Osanloo, 2014), the review begins by presenting the theoretical foundations of risk adopted in this study. Part 2.3 explores the technical aspects of AI and how they pose complex and emergent risks for AI for adopting organisations. The subsequent part explores the field of RM at both the process and organisational level. Drawing from these conceptual foundations, the review culminates with a critical depiction of the current state of AI RM literature, illuminating work at the frontier of academia to uncover the gaps which generate the research questions.

### 2.1 The Nature of Risk

Risk is a pertinent phenomenon faced by all organisations with the study of risk spanning multiple disciplines from psychology to mathematics to business. Conceptual and practical definitions of risk vary, and a unified definition of risk is unlikely. This work understands risk as resulting from the impact of an uncertain event on achieving business goals (ISO, 2009; Aven, 2013, 2016). This omits the broader philosophical conceptualisations of risk, along with their mathematically derived counterparts which exist within the risk nomenclature (Kaplan and Garrick, 1981; Aven and Renn, 2009; Andretta, 2014). This definition provides an understanding of risk that is both generalised and practically focussed, suited to the study of risks and their management from an organisationally centric perspective.

Organisational risk is multifaceted with expansive academic and practical literature attempting to form typologies to categorise risk in its multivariate forms. The result is a myriad of non-exhaustive categories, broadly grouping risks by their origin, characteristics, severity or impacts. Whilst a vast proportion of risk literature focuses on operational, financial and hazard risks (Razali and Tahir, 2011), various additional categories of risk are offered depending on contextual or industrial focus such as reputational, environmental, cyber-security, legal and supply-chain risk (Gaudenzi, Confente and Christopher, 2015; Mishchenko et al., 2021, Blundo et al., 2021). This is also true in the case of the finance sector where industrial conditions lead to various specific types of risk (Leo, Sharma and Maddulety, 2019). Kanchu and Kumar (2013) loosely categorise these into financial and non-financial (Figure 1). Ultimately, risk types are highly interdependent, with risks having knock-on impacts on other risks. Understanding these taxonomies of risk illuminates the areas in which AI systems can bring emergent risks to organisations.

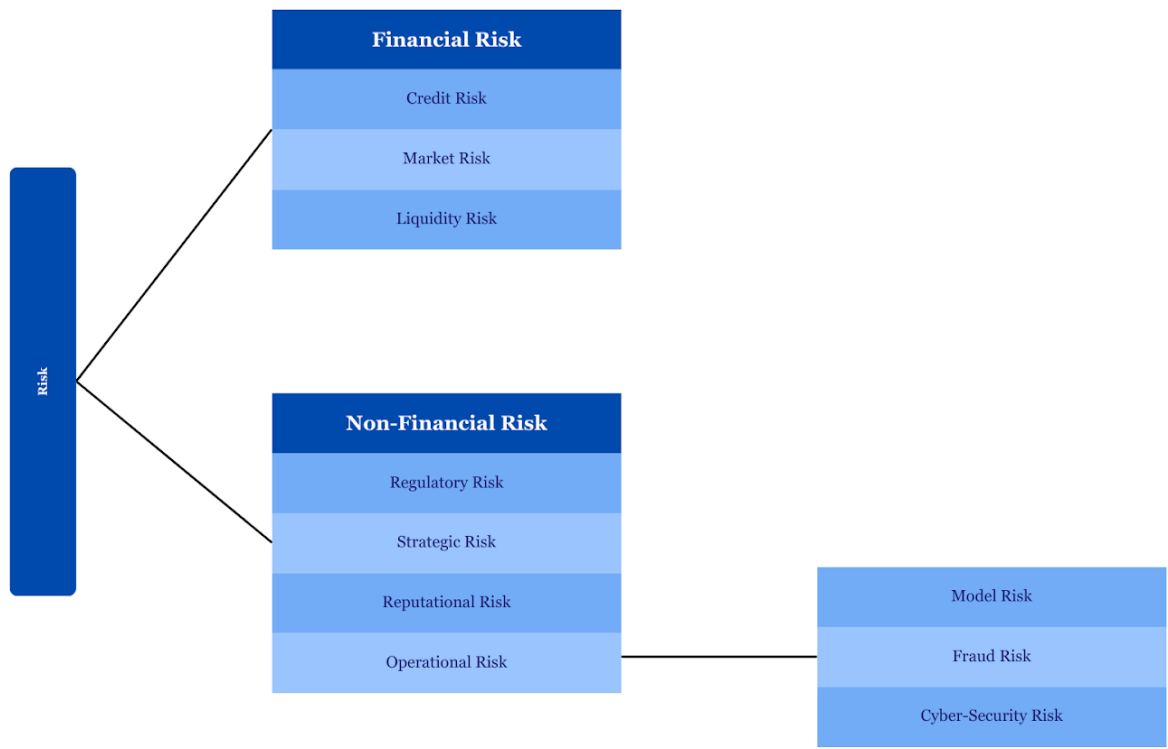

*Figure 1: Risk Types in the FS (Modified from source: Leo, Sharma and Maddulety, 2019, p.4)*

**2.2 AI and Organisational Risk**

With vast novel applications, and the ability to reshape organisational operations and decision-making, the risks from AI are unique, complex and pervasive. IT innovations are bound to bring evolving risks to organisations (Samimi, 2020). Yet in comparison with other IT systems, AI systems can be more dynamic, less transparent and can produce unintended consequences (Eitel-Porter, 2021). Uncertain or ambiguous, the risks that AI brings can be classified as emergent (Mazri, 2017). Since AI risks originate from the technical specificities of the AI systems themselves, understanding these technical risks accentuates their potential to propagate into organisational risks.

In what is argued to be the first systematic review of these technical risks, Zhang and colleagues (2022) identify two classes of AI system risk: data-level risk and model-level risk. Data-level risk arises as AI models are trained on vast quantities of existing data, learning from this and gaining the ability to make decisions and produce outputs. Whilst obtaining or holding data creates privacy and cyber-security risks, poor quality data can result in biased or inaccurate outputs (Mehrabi et al., 2022). On the other hand, model-level risk originates from the mechanics of the AI systems themselves. Whilst also being prone to issues like bias, a key model risk issue is transparency (Larsson and Heintz, 2020). Volatile and noisy datasets leave models more prone to errors, which is prevalent in the case of financial data (Ashta and Hermann, 2021).

The scope of the review by Zhang et al. (2022) focuses primarily on the first-order technical risks of AI systems, neglecting the wider qualitative ethical risks that these systems can pose. A vast proportion of literature accentuates these ethical issues. These are captured in an alternative taxonomy by Steimers and Schneider (2022, p.9), delineating AI risks between 'ethical aspects' and 'reliability and robustness'. Alongside ethics, fairness, accountability and transparency are key recurring themes within AI risk literature (Bogina et al., 2022).

On the organisational level, risks from AI models primarily constitute a form of model risk and operational risk (Garro, 2019). However, AI systems can present an array of ramifications on various types of organisational risk. For instance, unethical behaviour impacts presents a reputational risk but can also manifest into a financial risk (Fombrun and Foss, 2004). In a financial context, Boukherouaa and Shabsigh (2021) present five sources of

AI risk as bias, explainability and complexity, cybersecurity, privacy and robustness. Alternatively, Buckley et al. (2020) highlight regulatory and reputational AI risks as critical within finance firms. The reality for organisations is that AI risks are complex, interconnected and situationally dependent.

**2.3 Managing Risk**

There is a general literary consensus that organisations must manage risk to promote organisational competitiveness, stability and success (Elahi, 2013; Stein and Weidermann, 2016). As a result, the field of RM has garnered substantial scholarly attention over the previous four decades (Aven, 2016). RM can be broadly defined as the process of recognising and addressing risks in the effort to achieve business objectives (NIST, 2012). This definition embodies the inherent tension that exists within RM as risks intrinsically contain rewards and opportunities.

In its broadest sense, two paradigmatic approaches exist to RM. Proactive RM aims to recognise and address risks in advance, whereas reactive RM seeks to deal with risks as they materialise (Grötsch, Blome and Schleper, 2013). Despite intuitive arguments for the benefit of a proactive approach (Siegel, 2018), reactive RM practices grow in importance in environments characterised by uncertainty (Chapman and Ward, 2003; Marchant and Stevens, 2017). Thus, scholars see both approaches as complementary, arguing for a pragmatic and situational balance (Pavlak, 2004).

Emerging technological risks present unique challenges for incumbent RM frameworks (Isigonis et al., 2020; Samimi, 2020), and can expose existing approaches at all levels (Smith and Fischbacher, 2009). In uncertain environments, arguments are made for adaptive RM, enabling the alteration of RM frameworks as challenges emerge (Holling,

1978; Walker, Marchau and Swanson, 2010). Bjerga and Aven (2015) describe this as an iterative, collaborative and learning intensive approach to RM.

**2.3.1 Risk Management: Process Level**

Whilst academia is disjointed with aspects of RM, consistency exists over the generic activities involved in RM: risk identification, risk assessment, risk response and risk monitoring (Bandyopadhyay, Mykytyn and Mykytyn, 2002; Oehmen et al., 2020). However, this linear set of steps has been criticised for its relatively static approach to RM. Literature importantly stresses the cyclical nature of these processes where there is a continuous repetition of these activities as the risk environment evolves, leading to the conceptualisation of the RM cycle (Paltrinieri et al., 2014). This similarly forms the basis for much of the practical literature on RM where a plethora of principles and frameworks exist, notably COSO (2004), IRGC (2005), ISO (2009) and NIST (2012). Despite more specialised frameworks being offered, they still often lack the nuance to address the contextual granularity of RM in practice (Cedergren and Tehler, 2014).

Once risks are identified and assessed, organisations then determine a risk response. Four prevailing risk response strategies are evident in the literature. Often captured through varying terminology, these consist of avoidance, mitigation, transfer or acceptance (Bogodistov and Wohlgemuth, 2017). Two of these approaches are relevant within this context. The first is risk avoidance, which aims for elimination of the risk by evading the activity that causes it. The second is risk mitigation, in which processes and mechanisms are enacted to manage the risk entity over time (Tummala and Schoenherr, 2011).

**2.3.2 Risk Management: Organisational Level**

The operational focus of the RM cycle, along with its derivative principles, practices and frameworks, often lack the appreciation of the organisational level factors that occur during the process of managing risks. Risk governance emerged later as a complimentary stream of literature to RM. An influential paper by Aslet and Renn (2011, p.443) defined it as the 'critical study of complex, interacting networks in which choices and decisions are made around risks'. This intersects with the maturing field of Enterprise Risk Management (ERM) which advocates for the holistic and comprehensive management of organisational risks as opposed to traditional isolated approaches (Anton and Nucu, 2020). Achieving this in practice often requires teams of individuals independently dedicated to the management of risk on an organisational level (Hoyt and Leibenberg, 2003), often enacted through directives and audits (Kouns and Minoli, 2011).

Effective RM at the organisational level involves many aspects and is seen as integral in the management of AI risks (Dwivedi et al., 2019). A key consideration at this level is the structures through which RM takes place, where RM duties cascade down from the management level to the process level (Fraser and Henry, 2007). As Sheedy and Griffin (2016) note, the existence of these structures is insufficient to enable effective RM, and they are optimised by prudent risk culture and interaction. In an empirical analysis, Brookfield et al. (2014) confirm coherent communication as a crucial element of IT project RM. Insightfully, Nielson, Kleffner and Lee (2005) argue the importance of external alongside internal communication as central to effective RM. Overall, the combination of these aforementioned organisational aspects with their process focussed counterparts (2.4.1) constitutes an organisation's risk management framework (RMF).

## 2.4 The Contemporary Literature: AI Risk Management in Organisations
### 2.4.1 AI Risk Management in Organisations

Stemming from their emergent and pervasive nature, AI risks pose challenges for existing organisational RMF's. Steimers and Schneider (2022) contend that existing RM processes for software are unprepared to mitigate AI risks with Kruse, Wunderlich and Beck (2019) reporting this inadequacy in a financial context. Consequently, organisations are being exposed to a greater level and variety of operational and model risks, yet also the potential for regulatory, reputational, and financial risk among others. In response to this, Lee, Floridi and Denev (2020) argue for a greater inclusion of these non-model risks within organisations RMF's.

Whilst some scholars suggest the utilisation of traditional RM approaches in the face of AI risks (Clarke, 2019), a number of enhanced or novel approaches are evident in the literature. Due to the emergent nature and uncertainty of AI risks, certain paradigmatic approaches to addressing risk based on quantification and anticipation are inherently flawed. The quantitative approaches that have been offered often have limited applicability outside of narrow contexts (Bosnic and Kononenko, 2009; Fang, Dutta and Datta, 2014; Rabanser, Gunnemann and Lipton, 2019). In light of this, Budish (2021) argues for a more qualitative and responsive approach to AI RM, stressing the need for stakeholder inclusivity to combat AI's dynamism and situational variation. In a similar vein, Kruse, Wunderlich and Beck (2019) argue that AI RM should be agile and adaptable.

Despite lamentation over the lack of robust guidelines for AI RM, practical frameworks are beginning to emerge (Steimers and Schneider, 2022). The notable release of NIST's AI Risk Management Framework (2023) succeeds in providing a comprehensive template for the extent of AI applications. Yet, comparable to other generalised RM frameworks, it is criticised as lacking the nuance to match the highly contextual nature of AI. Thus it acts more as a set of guiding principles as opposed to a robust practically focussed RM enabler (Geelal et al., 2023).

In academia, various AI RM approaches have been proposed and empirically investigated. Broadly applicable scorecards are emerging, alongside tools and practices for algorithmic audits (Rismani et al., 2023). Conceptual AI RM work predominates and existing industry focussed empirical examinations of AI RM are limited. Broader empirical studies exist such as Rismani and other's (2023) analysis of ethical RM practices and Solomon and Davis's (2023) cross-industrial study of AI risk governance in Australia. Despite the latter work finding overall unpreparedness, it lacks practically driven remedies.

A common theme within the literature highlights the importance of managing risk at every stage of the AI development lifecycle (Geelal, 2023). After highlighting 21 challenges of AI RM in financial organisations, Kurshan, Shen and Chen (2020: 2) propose a 'system-level approach' to the management of AI model risk. Their approach underscores the importance of continuous risk monitoring at every level of AI design, development and operation. Despite its overt model focus, the key strength in their approach comes from the framework's modularity and customizability, allowing it to be applied to a range of AI use cases.

Another recurring theme is the need for human oversight of AI systems. Due to their inherent intelligence, AI systems can operate with autonomy. However in its current form, many AI systems provide an augmentative role, with human operators overseeing systems to identify erroneous outputs (Candrian and Scherer, 2022). Human oversight can exist in the form of periodical output audits, or a human-in-the-loop (HITL) integrated into AI system training and operation (Zanzotto, 2019). The maintenance of this division of labour between AI and humans is seen as fundamental to mitigate potential risks (Ashta and Herrmann, 2021).

**2.4.2 AI Risk Management in Financial Organisations**

Due to the manner of its organisational practices, the FS experiences a unique risk environment (Leo, Sharma and Maddulety, 2019). The intense regulatory landscape is in a constant state of flux, especially in the face of emerging technologies (Ducas and Wilner, 2017), heavily impacting the implementation of RMF's (Guidici, 2018). Overall, RMF's in the FS are especially mature (Christofferson, 2012), with extensive implementation of Model Risk Management (MRM), the process of screening and controlling risks in models (Kurshan, Shen and Chen, 2020). Nevertheless, coherent guidelines for AI RM in the FS are lacking (Lee, Floridi and Denev, 2021), and debates still exist over the preparedness of FS RMF's in the face of emergent AI risks.

The increasing role of models in shaping organisational decision making is increasing model risk (Cosma, Rimo and Torluccio, 2023). Growing criticism of MRM finds it overly detached from comprehensive organisational RM structures (Scott, Stiles and Debata, 2022). Complex and opaque AI systems can further jeopardise incumbent MRM frameworks (Gan et al., 2021). According to Brockte (2020), pressure on these frameworks builds further as AI systems are applied to unconventional areas in which existing MRM techniques are not well developed. Insightfully, Souza (2023) contends that existing MRM practices provide robust foundations to combat AI risk, yet need to develop risk identification, data-management and testing. Comparably on a wider scale, Lee, Floridi and Denev (2021) argue that the foundations of RMF's in FS organisations are reasonably equipped for the challenges of AI yet require particular alterations as risks emerge. The exact changes are in debate with some scholars arguing the importance of AI specific risk personnel to facilitate these adaptations (Schafer et al., 2022).

**2.5 Gaps and Research Questions**

Despite the number of conceptual works addressing AI RM, there is limited empirical literature on the topic, especially in the finance sector in which AI applications are relatively mature. The lack of understanding AI RM in practice has left academics calling for the need to expand this empirical body (Wirtz, Weyerer and Kehl, 2022). Alongside the lack of understanding of AI RMF's in general, the level of their preparedness in the FS is in question. Furthermore, the variety of existing disparate conceptual work has been argued to provide confusing guidance (Elliot et al., 2021), necessitating grounded empirical insight of best practices. Thus, the remainder of this study is based around the following research questions:

**RQ1**: Are FS organisations' existing risk management frameworks equipped for the risks of AI?

**RQ2**: How are FS organisations approaching the risks of AI in the context of risk management?

**RQ3**: What are the primary activities and mechanisms utilised by FS organisations for AI risk management on both a processual and organisational level?

**RQ4**: Are there any dominant principles, processes or mechanisms that are employed in AI risk management which may constitute a form of best practice?

# 3. Theoretical Foundations

In order to investigate these research questions and subsequently dissect the results, two key theoretical foundations are utilised.

## 3.1 Organisational Risk Management Framework

From the literature review, it is evident that RM is complex and that organisational RM requires a simultaneous utilisation of both RM processes and wider organisational regimes. As noted, the RM cycle is overly process focussed, and often neglects the wider mechanisms of RM at the organisational level. To capture both of these aspects, the RM cycle and the associated organisational wider aspects of risk governance are integrated by the author into a singular conceptual model (Figure 2).

Inspired by the notable integrated ERM framework presented by COSO (2004) and from Indrajaja et al. (2020), Figure 2 presents the fundamental activities of RM as situated within the wider structures and mechanisms of an organisational context. Therefore, these activities overlap organisational structures enabled through interaction. Drawing from the literature review, interaction encapsulates the mandates, audits, communication and cultural aspects that enable RM to occur (Spira and Page, 2003; Nielson, Kleffner and Lee, 2005; Fraser and Henry, 2007). The framework adopted here was guided by the simplistic yet fundamental elements of processual and organisational RM, providing a broad yet holistic lens to systematically conduct an empirical investigation of RMF's.

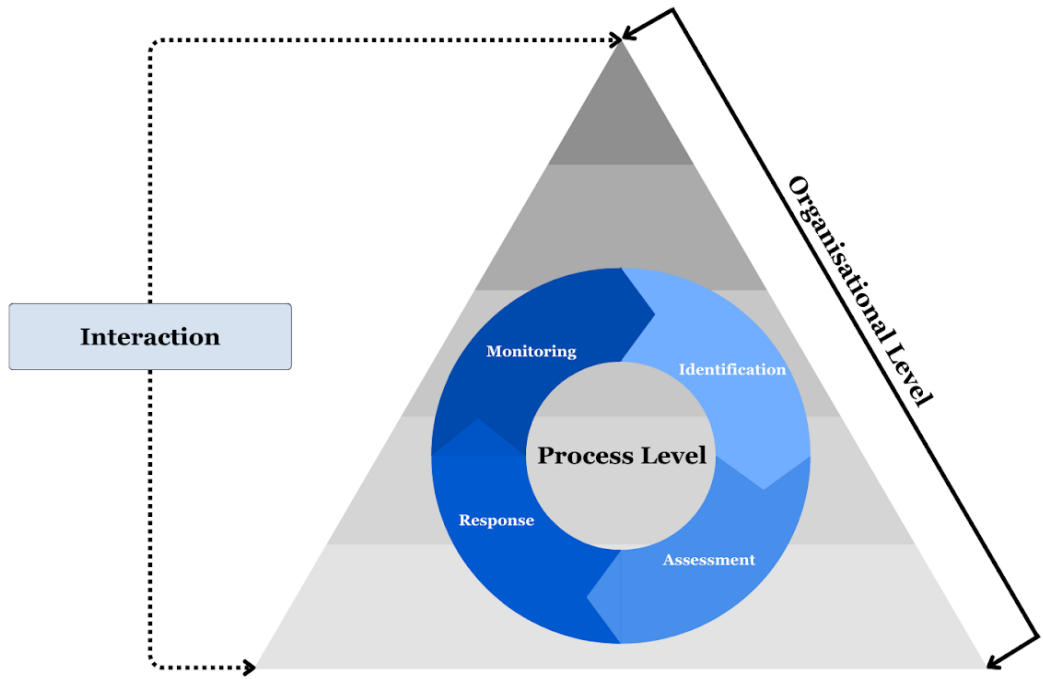

*Figure 2: The Organisational Risk Management Framework (Modified from source: COSO (2004); Indradjaja et al., 2020)*

### 3.2 Institutional Theory and Contingency Theory

A blended lens of contingency and institutional theory can help dissect convergent and divergent approaches to AI RM within organisations. Contingency theory is one of the predominant theoretical lenses drawn upon to understand the architecture of organisations (Donaldson, 2003). It posits that there is no best practice of establishing control systems such as RM frameworks, and thus observed frameworks are context contingent (Otley, 2014). Whilst generally applicable, scholars have found its utility within RM research (Hanisch and Wald, 2012). Context contingency is often attributed to a number of variables, with RMF contingency said to rest upon the nature of the risk and circumstances of the organisation (Mikes and Kaplan, 2013). Leveraging this theory, it can be hypothesised that organisations' AI RMF's will exhibit a level of divergence as they are tailored to their internal and external conditions, and the context of AI risk.

On the other hand, institutional theory proposes organisational RM consistencies in response to their social, cultural and regulatory environment (Zsidisin, Melnyak and Regatz, 2005). Regulations are one of the key institutional drivers of RM implementation (Collier et al., 2006; Filatotchev, Jackson and Nakajima, 2013). Together with public legitimacy considerations and recognised guidelines, they reflect the formal and informal forces that drive coercive isomorphism (DiMaggio and Powell, 1983; Husin and Oktaresa, 2011). Alternatively, convergence can occur through mimetic isomorphism, in which organisations replicate one another due to uncertainty over the correct way to act (Hudin and Hamid, 2014). Advocated by Sarens and Christopher (2005) as a robust lens to study organisational RM, the literature review suggests the possibility of both forms of isomorphism being influential within the context of this study.

Fundamentally, contingency and institutional theory are at inherent odds. In isolation, both theories are limited due to their reductivity. Whilst contingency theory is overtly generic (Donaldson, 2006), institutional theory is criticised for framing organisations as too passive (Scott, 2008). Yet combined, these theories can help explain both the homo- and hetero- geneity of organisational RMF's and their approaches to risk. This blended approach has been utilised in previous work in attempts to understand the mechanics behind RM phenomena and permit observable comparisons between RMF's (Suardini et al., 2011; Hudin and Hamid; 2014).

## 4. Methodology

Emerging from the literature, Chapter 2 uncovered research gaps and permitted the formation of four research questions. Following, this chapter outlines and justifies the research methodology adopted in this paper in the face of these research questions. In the field of qualitative research, open and systematic disclosure of research methods, and the processes and logic underlying them is imperative to offer robust and rigorous qualitative research (Nowell et al., 2017). The chapter begins by advocating for the qualitative methodology employed in this study, detailing the data collection and analysis methods employed as a result. The subsequent parts of this chapter provide a critical interrogation of this study's methods and address the measures enacted to ensure its ethical conduct.

## 4.1 Research Design

### 4.1.1 Methodology

Due to this study's exploratory nature and the emergent and varying qualities of AI risk management, a qualitative interview-based approach was deemed to be the most suitable method of data collection. This will allow rich insight into RM approaches within finance sector organisations. Despite the value of utilising a quantitative methodology, the exploratory nature, and difficulty accessing participants in the quantities needed to produce robust quantitative results drove the author to follow an approach focussed on depth as opposed to industry wide generalisations. In support of this, Myers (2009) and Eriksson and Kovalainen (2015) advocate for qualitative methods as a robust standalone approach within organisational research, despite its longstanding tradition of quantitative methods.

### 4.1.2. Justification of Methods

Despite the potential of various qualitative methods, in-depth interviews were selected for this study. Whilst focus groups and case studies were both suitable and capable of producing high quality data, certain constraints made them unviable. Focus groups present critical confidentiality issues (Sim and Waterfield, 2019), contradicting the strict anonymity requested by participants in the case of sensitive corporate information. Meanwhile, due to participant access constraints, a major challenge of case study research would be the potential to obtain enough rich data from multiple companies to provide sufficient generalisability and validity (Glette and Wiig, 2022).

The interviews followed a semi-structured framework. This crucially provided flexibility, a fundamental doctrine in the practice of exploratory research (Stebbins, 2001), and a necessity given the lack of unified literature on the topic. Beneficially, this approach affords sensitivity toward emerging topics, allowing them to be probed in greater depth. Despite this, the initial part of the interview framework was kept rigid, with consistent questions across the interviews providing the basis for systematic comparison, whilst minimising bias originating from the role of the interviewing researcher (King, 2004). This was beneficial to account for the variation in the manifestations of AI between respondents expected from the literature review, and provide an in-depth and thorough understanding (Carruthers, 1990).

In support of the methodological choices outlined in this section, various other studies adopt similar qualitative approaches to study RM within organisations (Ali and Naysary, 2014; Hohma et al., 2019; Nasteckiene, 2021). In a review of RM practices in small and medium-size enterprises, around half of the empirical papers reviewed were qualitative, and a third of those relied solely on interviews (Falkner and Heibl, 2015). Wood and Ellis (2003) used standalone semi-structured interviews to determine the RMF's adopted by a sample of UK cost consultants. Meanwhile, Rismani and colleagues (2023) use interviews to conduct exploratory research on organisational AI RM practices.

## 4.2 Data Collection

### 4.2.1 Participant Selection

Purposive sampling involves the selection of participants who possess certain qualities or experiences (Etikan et al., 2016), and is utilised to identify 'information-rich cases' who can provide valuable and relevant insights (Palinkas et al., 2015, p.553). Employing this method, participants were selected based on their employment within the finance sector, and robust knowledge of both AI and RM practices within their respective organisations. Thus despite interviewing a singular participant from each organisation, the gathered data on RM can be extrapolated to the organisational level, reflecting their organisation's overall approach to AI risk management with reasonable certainty.

According to Patton (2014, p.264), qualitative research tends to involve small samples scrutinised in depth. Thus, it is imperative that a sample is selected pragmatically to capture nuanced variations within the study sample (King, 2004). To provide both an understanding of the wider industry, as well as a holistic picture of RM within organisations, effort was made to draw participants from a range of different types of financial organisations and roles within them. To achieve this, contact was made either through direct email to publicly accessible addresses, or through email addresses obtained through professional contacts within the finance sector. To maximise potential involvement from an inaccessible population, Salganik and Heckathorn (2004) argue the benefit of snowball sampling. This technique was employed, yielding 3 extra participants.

### 4.2.2 Interview Conduct

The study involved conducting interviews with 9 individuals from different financial organisations lasting between 36 and 75 minutes and taking place between June and August

2023. A rigorous interview guide is essential to ensure the quality, objectivity and plausibility of interview-based studies (Kallio et al., 2016). To ensure this, the development of the guide followed the 5 step process presented by Kallio et al. (2016), notably including a pilot interview to assess the guide's suitability. To maximise cooperation and build rapport, interviews followed an inverted funnel method (Mandel, 1974). This began with an explanation of the meaning of RM adopted in this study and broad questions surrounding participants' roles, AI applications and AI risks within their organisations. Aligned with the research questions, the remainder of the interview was driven around a number of more focussed questions concerning their organisation's AI RM practices. Whilst the basic structure of the interview was kept consistent across all participants to facilitate inter-organisational comparison, nuanced alterations were made fluidly as interviews took place. Aligned with the advice of King (2004), this permitted the exploration of emerging lines of enquiry.

**4.3 Data Analysis**

Data collection and analysis were undertaken simultaneously to facilitate iterative and reflexive data analysis (Srivastava and Hopwood, 2009), and provide flexibility to refine interview templates as the project was conducted. In this study, data was analysed through inductive thematic analysis: a method used to identify, analyse and produce themes which represent patterns within a dataset (Maguire and Delahunt, 2017). Consistent with the study's exploratory nature, the analysis was undertaken inductively, following Braun and Clarke's (2006; 2020) six step process. The first step, familiarisation, was partly enacted through transcribing interviews by hand (Byrne, 2022). As the level of specificity should match the required depth required by the research objectives (Bailey, 2008), transcriptions were made verbatim. Participants were also provided with transcripts to report inaccuracies in an attempt to maximise the study's validity.

Using NVivo14 (Lumivero, 2023), open coding was then conducted, with attempts to remain objectively detached from preconceptions around content meaning (Cascio et al., 2019). Viewing these codes within the context of the research questions allowed the generation of four key themes. According to Graneheim, Lindgren and Lundman (2017), qualitative research becomes more credible with less abstracted and interpreted themes. Hence, themes were derived from the literal processes and activities implied by the collection of underlying codes. Through this process, the author kept note of the particular use case of AI adopted by the organisation to contextualise their approach. Emergent themes were subsequently refined through iteratively assessing their representation of the data collected.

## 4.4 Methodological Limitations

The researcher wishes to note a few core limitations of this research methodology. Primarily, due to time constraints of completing the project, and corporate hesitance due to the potentially sensitive nature of the topic, the sample size was constrained. Whilst holding potential for sample bias, the scope and complexity of the industry studied makes it likely that saturation was not achieved (Glaser and Strauss, 1967). Since various scholars champion saturation as a fundamental element of robust qualitative work (Fusch and Ness, 2015; Morse, 2015), the lack thereof is likely to make the conclusions from this research less generalisable and reduce its external validity. However, the population sampled did exhibit heterogeneity and by the final interviews fewer novel concepts were being unearthed. Moreover, the main aim of the project was to gain in-depth empirical insight of AI RM within finance companies, as opposed to deriving an industry-wide insight of the contemporary state of the phenomena.

A second key limitation arose from the uncertain accuracy of participant responses giving an inaccurate or incomplete picture of RM practices within the surveyed organisations. Firstly, due to the corporate sensitivity of the research topic, certain topics may have been avoided, although trust building and anonymity assurances likely reduced this. Secondly, as no individuals' roles were within organisational risk teams, they may have imperfect knowledge of AI risk management. Despite this, participants selected possessed reasonable knowledge of their organisation's RM approach, thus this is likely to have a lesser impact on this study's validity.

Finally, it is imperative to note the subjective aspect of qualitative research in which the researchers' preconceptions can influence the inherent nature of the study. This draws the need for reflexivity, which is the critical examination of the 'role of the self in the creation of knowledge' (Berger, 2015, p.220), with the ability to improve credibility of qualitative research. As an inexperienced researcher, expectations of RM realities influenced the process of data collection, distorting the direction of interviews. Aware of this, the author remained critical of their position through periodically reflecting upon how their expectations compared with the data. Through keeping questions broad, iteratively adapting the interview guides, and coding based on literal concepts, the author believes a more objective account of the phenomenon was obtained.

## 5. Findings

The following sector presents the findings of this study. Inductive thematic analysis has identified four central themes aligned with the central research questions. Theme 1 is mainly concerned with RQ1 and RQ2, Theme 2 with RQ2, and Themes 3 and 4 with RQ3. The amalgamation of these findings then constitutes the basis of RQ4, examined within the discussion. The section begins with a summary of the interview participants and their organisations, detailing participant roles and their organisation's AI applications to contextualise these findings.

**Table 1: Participant and Organisation Characteristics**

| Firm | Area of Finance | Participant | Role | Number of Employees | AI Applications |
|---|---|---|---|---|---|
| A | Asset Management | A1 | Regional Manager | 100+ | None |
| B | Asset Management | B1 | Developer | 100+ | Predictive Modelling |
| C | Wealth Management | C1 | Developer | 100+ | Predictive Modelling + Asset Allocation |
| D | Asset Management | D1 | Investment Manager | 100+ | Predictive Modelling + Portfolio Optimization |
| E | Asset Management | E1 | Digital Enablement Director | 1000+ | Backend Optimisation |
| F | Investment Bank | F1 | Product Manager | 1000+ | Portfolio Management, Client Relationship Management, Trading Strategies |
| G | Financial Planning | G1 | Director | 1000+ | Auditing Financial Advice, Client Analytics, Employee Training |
| H | Financial Advisory | H1 | Senior Manager | 1000+ | Backend Optimisation, Call Summarization, Client Analytics, Robo Advice |
| I | Hedge Fund | I1 | CEO | 1000+ | Data Management, Onboarding, Trading Strategies |

## 5.1 Theme 1: Vulnerabilities: Framework Maintenance or Adaptation

The participants conveyed a complex landscape of AI risks, with regulatory, data, and system failure risks being pertinent for all participants. Those who were leveraging AI for predictive modelling (B, C, D, E) were most aware of model risks in the form of errors or system failures, conflating these with potential financial risks. Participants from firms F, G, H and I, who's companies had implemented more extensive systems which often had impacts on stakeholders and clients, were more aware of regulatory and reputational risks. All participants were aware of the challenges surrounding data accessing and availability, including privacy and cybersecurity concerns. Interestingly, B1, E1 and G1 noted systemic risks emerging from organisations overreliance on AI models capable of failure. Overall, all participants reflected how AI has the potential to challenge RM approaches, and that there was no best practice to guide these processes.

### 5.1.1 Subtheme 1: Maintenance

Four organisation's RMF's were well equipped for the risks of AI. For these firms (B, C, D, E), referred to as the *Maintainers*, the implementation of AI did not drive the imposition of any novel practices, activities or mechanisms to counter the related risks. The Maintainers integrated AI into their existing models, acting to optimise these systems. This is opposed to AI being used for novel purposes, fundamentally altering organisational processes and decision making, and thus its risk landscape.

Three of these firms (B, C, D) were using AI to enhance the predictive capacity of their models for investment strategies. In this capacity, the predominant risk discussed was the risk of errors, complexified by the opacity of AI models. Pre-deployment, existing MRM practices consisted of extensive testing in order to assess model predictability and robustness.

In operation, MRM consisted of human oversight of model outputs, in the form of a HITL or audits, facilitated *risk monitoring*, and acted as a line of defence before model outputs were translated into actionable decisions. Participants reflected how AI integration had a minimal impact on firms overall risk landscape, and existing MRM was competent in managing these risks:

*Black box models have been going for 30 years and at the end of the day the regulatory licence holder is responsible for the black box [...] auto-checking and human overrides have always been in place, and still are with AI.* (B1)

*It doesn't matter if a trade is generated by a human or a machine, ultimately the override catches it as it happens, so that's not a big worry for me.* (D1)

The outlier was Firm E, which used a custom-built AI tool from Microsoft for backend process automation. This firm's organisational RM practices were already equipped for AI driven automation after adapting them to the risks of previous non-AI automation systems. However, as they were in the low-level stages of implementation and using an assured prebuilt tool, the participant related how the risk landscape with and without AI was effectively unchanged.

*We've gone through that process [adaptation of risk management] when robotic process automation came in, so those risks were already part of our process design.* (E1)

Despite this, all participants from Maintainers noted the strain that would be caused to their RM practices if AI systems were given more autonomy, or applied to novel areas, especially in a client facing capacity.

**5.1.2 Subtheme 2: Adaptation**

Four other firms' RM practices were partially unprepared for the risks of AI. These made varying degrees of adaptation to their RM practices, activities and mechanisms in the face of AI implementation, and are referred to as *Adaptors* (G, H, I, J). For these firms, AI was utilised to develop novel processes, presenting emergent properties and an evolved risk landscape. Whilst participants believed that their existing RMF's were mostly equipped for the risks of AI, they built upon their frameworks in areas in which AI exposed vulnerabilities. Adapting firms saw the development of RM principles as a gradual process, choosing to develop their approaches as implementation cases expanded and novel risks were experienced. As F1 remarked:

*We know that traditional frameworks cannot be applied and that's why we're not forcefitting it [...] we would like to evolve our [risk management] framework and customise it for our own when we actually come to that point.*

For all of the Adaptors, human oversight, in the form of a HITL or auditors, was restructured toward, or integrated into, all novel AI applications to enable risk response and monitoring. Organisational learning regimes were improved in these firms to better enable these individuals to mitigate AI risks. Meanwhile, all Adaptors implemented contingency plans in case of system failure. Within this sample of firms, a range of other adaptations to RM and governance were observed. These included increasing the occurrence and AI focus of risk audits to match the speed of technological evolution (H, I), creating cross skilled teams of AI developers and risk experts (G, I), implementing protocols and software to enable greater explainability of AI models (F, I), and altering data handling policy to account for increasing sensitivity of model data (H).

Despite these alterations, participants from Adaptors relayed how certain aspects of their existing RM approaches were equipped for AI. Generally, approaches to data handling were robust, with data being kept internally, which participant's attributed to the stringent regulatory environment. Meanwhile, whilst the majority of adaptations related to risk response and monitoring, risk identification and assessment practices were largely unchanged. Participants reflected how AI risk assessment practices were undeveloped with an absence of quantifiable metrics.

## 5.2 Theme 2: Approaching AI Risks and AI Risk Management

This theme dissects three central aspects that define the way in which participants approached AI risks within AI RM. Although non-exhaustive, these aspects exhibited the highest degree of commonality across the participants surveyed.

### 5.2.1. Subtheme 1: Avoidance

When faced with risks that exceed a firm's appetite, an organisation's *risk response* may be to avoid those risks entirely. In response to the potential risks of AI, firm A elected to avoid these risks, opting to maintain their current non-AI models by virtue of their transparency and effectiveness. Despite the other firms choosing implementing AI in certain areas, risk avoidance toward particular systems was still evident. All firms were reluctant to implement autonomous AI systems, and systems whose outputs were not protected from impacting upon clients by human oversight. Participant G1 commented:

*Where we are trying to manage some of those general AI risks, I guess, is by putting none of this straight into the client domain.*

Risk avoidance was similarly professed as a result of particular AI risks such as opaque systems, and potential for regulatory and reputational repercussions. Therefore, AI deployment was gradual and predominantly across in low-risk cases.

**5.2.2 Subtheme 2: Reactivity**

A key principle for approaching RM under the case of AI drives the need for RM plans to provide rapid reactive responses to unforeseen risk manifestations. All participants were particularly aware of the emergent and unpredictable nature of AI risks. In response to this, participants F1, G1, H1 and I1 reflected how traditional proactive approaches to risk were put in jeopardy. For them, the answer required more robust reactive RM once risks have materialised. This included strengthening the procedures and regimes to allow a rapid analysis of the underpinnings and mechanics of the risk event, learning from this and improving RM strategies to counter existing inadequacies. As participant I1 argues:

*What I think that is often neglected is to have a really robust and timely lesson learned process rather than something that's open-ended and vague [...] there should be an understanding about what you have to put in place when incidents occur, whose job it is to deal with it, what the endpoint is.*

Participants also discussed the integration of contingency plans such as reverting to previous manual systems (G1, H1).

### 5.2.3 Subtheme 3: Responsiveness

Responsiveness to all AI risks as they evolve is a key multidimensional issue expressed by participants. All participants noted the rapid pace of contemporary AI development, highlighting the need for responsive approaches to AI risk and flexible RMF's as a result. One key dimension of this noted by the majority of participants is responding to AI risks throughout the AI development lifecycle. On the other hand, responsiveness can relate to risk *identification*, where emergent risks must be promptly identified (B1, E1, G1). Developing on this, participant I1 stressed the need to incorporate less likely, outlier risks into risk *identification* and *assessment* practices, alongside the more salient ones.

A second critical issue relating to responsiveness was attributed to the lethargic and sporadically evolving regulatory landscape (A1, D1, E1, G1, H1). Participants stress how this requires foresight in risk *identification*, and rapid responses as the landscape alters (H1, G1). To counter the fast-evolving landscape of AI and regulation, participants reflected the need to conduct more regular committee meetings and audits to identify risk and reconfigure frameworks (H1, I1).

## 5.3 Theme 3: Risk Management Activities

From this study, two predominant processual activities were determined across all companies AI RM approaches: Human Oversight and Testing.

### 5.3.1 Subtheme 1: Human Oversight

Human oversight was described to be a defining and essential constant within all of the firm's RM approaches. Participants from both adaptors and maintainers relayed that their firms relied on either a HITL or human oversight audits of system outputs as a first line of defence against erroneous AI outputs. Reducing the autonomy of AI systems organisational decision making, all organisations integrated some form of oversight into all AI systems. This was described to provide continuous risk *monitoring* of systems to identify errors (B1, C1, D1, E1, F1, I1), or potentially harmful bias outputs (G1, H1). Human oversight was observed by all participants during the operational deployment of models, all participants deemed it necessary during the design, training and testing process. Overall, human involvement was seen as the most comprehensive strategy to reduce AI risk, keeping within accountability mandates and providing a bridge between AI outputs and organisational decision making to minimise the impact of erroneous outputs propagating through the business. Wider implications of human oversight were also reflected. This approach reduced the need for contingency plans (G1), and monitoring outputs was also argued to reduce the risk of opaque, complex models, and issues of poor quality data (B1, C1).

*Things like hallucinations and or profanity or bias all that kind of stuff, we've always got an advisor at some point in the process. You're kind of creating a layer there if you like.* (H1)

Human oversight was multi-layered, with model developers and product owners overseen by organisational risk oversight committees. In spite of this approach, multiple participants noted the flaw in human oversight as humans can be fooled by convincing, but incorrect, system outputs (D1, F1, H1, I1). Participants H1, G1, and I1 discussed the importance of increasing training to better equip human overseers with the capability to identify and manage AI risks.

### 5.3.2 Subtheme 2: Testing

Robust model testing also played a critical part in organisations RM, widely described as a necessity in the face of FS regulations. Participants described how extensive testing aimed to discern the predictability of models, especially in the face of opaque systems (B1, C1, E1, F1, I1). This becomes a particular issue as traditional auto checking struggles to operate on complex AI source code, requiring human checks (F1). Participants B1 and C1 described how their testing was a multilayer process, involving multiple individuals cross-testing models in different software. Testing was not just described to be necessary pre-deployment. Once deployed, models may need to be periodically retested, and recalibrated, as the data they are analysing evolves (B1, C1, D1, E1, F1, I1). When dealing with external AI tools, such as ChatGPT, participant H1 reflected how safe environments must be constructed to test these tools before they are exposed to internal data.

## 5.4 Theme 4: Governing Risk

Opposed to particular activities of managing risk, participants provided some key insights with which AI risks are managed at the organisational level.

### 5.4.1 Subtheme 1: Structures and Burdens

All participants mention the hierarchical and decentralised burden of RM within their organisations in which RM is a combined effort from risk oversight teams and those closest to the models: developers and product managers. For all companies, the primary burden of AI RM fell upon those closest to the models. E1 noted that these individuals *'have a vantage point across tech and business'*, providing a critical connection between the technical aspects of AI systems and the wider risk related organisational concerns. Participants described how these individuals have the capability to understand models and identify issues and risks. However, multiple participants highlighted the need for these individuals to be better trained in order to diagnose system issues (D1, F1, G1, H1, I1). Meanwhile, risk oversight teams, with an overarching perspective of organisational risk, provided two functions: to conduct oversight and risk audits as a final line of defence, and derive the overarching processes, procedures and regulations through which AI development and operation fits within. Discussing the relationship between these two actors F1 adds:

*Risk management is compartmentalised in that sense for those closest to the models. So we have a product owner and as that product develops we'll have steering codes that oversee that development and ask questions. We're basically trying to make sure we've got risk experts on those steer codes to ask the questions and to probe. There's no kind of centralised way of dealing with that yet.*

Participants highlighted how the balance and structure of RM burden over AI systems needs to be carefully defined in order to strike a balance between risk exposure and innovation (G1, H1).

### 5.4.2 Subtheme 2: Communication

Participants reflected how open, regular and transparent communication is fundamental between product managers and oversight committees (B1, C1, D1, F1, G1, I1). This can be achieved by implementing clear protocols (F1) and designing rapid communication channels if risks were to occur (G1, I1). It also involves an external dimension, as multiple participants involved third parties in the advancement of their RM principles (F1, G1, H1). A crucial communication issue was highlighted by participant C1 as complex models are difficult to describe to less knowledgeable oversight committees. In response, C1 elected to offer detailed explanations of model mechanics to facilitate understanding. In the context of coherent communication, participant I1 reflected how expertise must be shared between individuals who understand risk, AI and the processes behind the models respectively:

*You've got to platform AI capabilities in a cross-skilled team, where someone who understands AI is sat with someone who knows the purpose of the models and people from with the risk team working together on this stuff.*

## 6. Discussion

As artificial intelligence implementation grows, the fundamental capabilities of the technology are harnessed in tandem with the novel risks that it can bring. The intention of this study was to provide exploratory empirical insight into how organisations are approaching and managing these risks within the context of the FS.

With respect to RQ1, the study suggests that AI integrated into existing models falls within existing RMF's, whilst novel applications put strain on them. Meanwhile, when approaching AI risks (RQ2), organisations were shown to either adapt or maintain their existing frameworks dependent on the nature of AI risk realised through the integrated or novel capacity of AI systems. Approaches of avoidance were prevalent toward excessive risks of autonomous AI, with RMF's requiring reactive measures and multidimensional responsiveness in the face of emergent AI risks. Two prevailing processual activities, and two organisational level aspects of RM were identified aligning with RQ3. Drawn from these findings, some aspects of best practice are formed (RQ4). To frame and contextualise the findings, this discussion draws upon the nature of AI risks as emergent to unravel the approaches organisations have taken toward them. Meanwhile, institutional and contingency theory dissect organisational convergence and divergence.

## 6.1 Equipped and Evolving Frameworks in the Context of AI

RQ1 asked: are FS organisations existing RMF's equipped for the risks of AI? It is well noted that the novelty of AI risks has the potential to expose vulnerability in existing RM approaches (Kruse, Wunderlich and Beck, 2019; Zhang et al., 2022). The study suggests that, on the whole, the surveyed finance organisations RMF's are adequately positioned to manage the risks of AI. This is particularly the case when AI is integrated into existing models with developed MRM practices. Meanwhile, when AI systems drove novel applications, targeted adaptations were required to better equip these firms for the evolved risk landscape. This then led to two distinct approaches to AI risks and their management (RQ2): framework maintenance or framework adaptation.

### 6.1.1 Maintenance: Efficacy of Existing Frameworks

One of the primary debates identified within the literature related to the preparedness of existing RMF's in responding to the emergent risks of AI implementation. Gan et al. (2021) contend that well developed practices of MRM within the finance industry can be put under strain as models become increasingly complex and opaque with the introduction of AI. In spite of this, the evidence from the Maintainers suggests that the risks of lower-level AI integration into existing predictive models are well managed by existing MRM and their associated oversight regimes. In part, this is due to the lack of novelty that AI risks present, where opaque and complex systems were in existence prior to the integration of AI. Above all, human oversight within existing MRM acts as a robust line of defence between model outputs and subsequent organisational decision making. Yet, for these firms, AI was not leveraged to introduce new models and drive novel applications. As Brockte (2020) suggests, novel applications are more likely to expose vulnerabilities in MRM practices. This potential was highlighted by multiple participants, thus if firms surveyed had developed these more extensive applications, adapted MRM practices may have been observed.

Despite results from the outlier firm seeming to suggest that existing RM techniques for robotic-process automation are proficient with the introduction of AI, it is essential to note that the low-level nature of AI within this implementation and use of a prebuilt tool likely means these results would not hold for the extent of more sophisticated AI automation systems. As these more sophisticated systems can pose a number of exacerbated risks (Syed et al., 2020), they are more likely to necessitate RM adaptation.

### 6.1.2 Evolution: Framework Adaptation

Even for the Adaptors, various aspects of their existing RMF's were equipped for the risks of AI. However, when the implementation of AI drives the emergence of novel risks, organisations are likely to experience gaps in their RMF's (Marchant and Stevens, 2017). Scholars contend that reacting to emergent risks requires leveraging and adapting existing frameworks, as opposed to creating new ones (Almeida, Santos and Farias, 2021; Mokander et al., 2022). Aligned with this, the Adaptors noted the various ways in which their RMF's had been incrementally adapted in response to particular vulnerabilities. Somer and Thalman (2023) portray a similar picture of the practicalities of AI RM in an industrial context in which existing frameworks provide partial coverage of AI risks, and act as sufficient platforms on which to evolve RM.

The study evidenced that risk response and monitoring activities underwent the greatest transformation with human oversight regimes restructured and training increased to help users monitor and mitigate AI risks. Overall, the alterations made by the Adaptors allowed them to adjust to the nuanced risks posed by their AI systems, improving knowledge and altering procedures to better equip them to the dynamism and opacity of AI risks. Whilst this indicates that these organisations RMF's weren't fully equipped for the risks of AI, it does not go as far as indicating the logic behind these particular alterations.

**6.1.3 Maintenance and Adaptation as Approaches to AI Risk Management**

Maintenance and adaptation represent two contingent approaches to AI RM. The dichotomy between these two approaches was underlined by the nature of the AI systems employed. Whilst Maintainers AI systems were integrated, Adaptors systems were novel. Thus, the risk profiles of each were fundamentally different, with novel systems introducing evolved risks, impacting the approach taken. Contingency theory underlines this processual and architectural differentiation in RM approaches between organisations which vary based on internal and external factors (Donaldson 2006, Mikes and Kaplan, 2013). These findings develop on the work of Mikes and Kaplan (2013), who assert that risk types are a key contingent factor in the variation of RM approaches between organisations. Whilst traditional contingency theory research underlines various other variables which influence the contingent nature of organisations control systems, understanding their relevance within this context was beyond the scope of this study.

**6.2 The Nature of AI Risks and Approaches**

**6.2.1 Approaching AI Risks**

Theme 2 outlined three overarching approaches toward AI RM (RQ2). These align with the fundamental nature of AI risks and their conceptualisation as emergent risks. Uncertainty and ambiguity are inherent to emergent risks, and thus they challenge existing RM strategies and require a distinctive approach when compared to traditional risks (Brocal, Sebastian and Gonzalez, 2017). These findings thus echo much of the literature surrounding approaches to emergent technological risks, whilst extending their relevance in the context of AI.

### 6.2.2. Risk Avoidance

Avoidance is one of the fundamental risk responses. One participant chose to avoid AI implementation all together due to the balance of AI risks with the efficacy of their existing models. Watt and Wiley (2016) contend that proven incumbent technologies can be favoured in order to avoid the risks of implementing novel ones. A prevailing area of avoidance was toward autonomous and client facing AI systems. Autonomous systems can present their own array of specific risks (Zech, 2018), and exacerbate regulatory risks (Wong, 2020), and thus the hesitance toward implementing fully autonomous AI amongst participants is rational and shared across various industries and AI applications (Abramoff, Tobey and Char, 2020). Meanwhile, accountability is tightly regulated within the FS (Zetsche, 2022), with fully autonomous systems acting to compromise this. Thus, the avoidance of these risks was heavily attributed to regulatory pressures.

### 6.2.3 Responsiveness as a Multidimensional Issue

Novel technologies challenge incumbent approaches to risk identification and assessment (Escande, Proust and Coze, 2016). Constant horizon scanning of outlier risks has been advocated due to AI risk ambiguity (Anderljung et al., 2023). Thus, RM should become a dynamic and continuously evolving process as these risks unfold (Yuan, Tan and Li, 2008; Anderljung, 2023). This approach is similarly advocated in the context of evolving software projects (Dorofee et al., 1996). One dimension of responsiveness identified is related to the identification of all risks throughout the AI development lifecycle, supporting AI centric risk literature (Anderljung et al., 2023), as well as within general technology risk literature

(Brocal, Sebastian and Gonzalez, 2017). Emerging risks furthermore present fundamental issues for regulators as well as organisations. Wansley (2016) relates this to the epistemic issue of ambiguity around the impact of risks and efficacy of regulatory responses before risks have manifested. This is a pertinent issue within financial regulation, and the regulation of new technology (Moschella and Tsingou, 2014), leading to ex-post regulation, a significant consideration to which RMF's must be attuned.

### 6.2.4 From Proactivity to Reactivity

The findings here suggest the increasing development and strengthening reactive RM practices in the face of emergent risks. As these risks are uncertain, proactive strategies based on anticipation fail to adequately protect against risk events, thus the participants argued the necessity of clear processes and protocols in the event of unforeseen risks. This supports literature which calls for pragmatically leveraging proactive and reactive strategies (Pavlak, 2004), especially within the context of emergent risks (Marchant and Stevens, 2017). Whilst this study highlighted the importance of such an approach, it provided limited insight into the design of reactive strategies and the areas in which they are needed. Interestingly, Hohma et al. (2023) suggest that reactive approaches to AI risk must comprehensively involve both RM teams and those closest to the models. Irrespective of the potential benefits of reactive RM in the face of emerging risks, Qian, Fang and Gonzalez (2012) argue that organisation's over reliance on reactive strategies can present a potential risk in itself. Thus, finding a proactive versus reactive balance is a critical issue.

### 6.3 Dominant Processes and Structures

RQ3 intended to investigate AI RM aspects at both a processual and organisational level. This study has uncovered two predominant processual RM activities and two defining organisational level aspects within FS organisations in the context of AI. Whilst other processes and mechanisms were identified from participants, the following were the most relevant and widely used across all organisations surveyed.

**6.3.1 Human Oversight Regimes**

Amongst the wide range of potential RM approaches toward AI, this study suggests overwhelming implementation of human oversight regimes. Human oversight can exist throughout the training and operation of AI systems (Mosqueira-Rey et al., 2022). Above the potential to provide more powerful AI systems (Xin et al., 2018), a human-in-the-loop can work towards facilitating robustness, safety and compliance (Buckley et al., 2021). It allows for constant monitoring and also alleviates accountability debates when autonomous AI systems are utilised for decision making (Etzioni and Etzioni, 2016).

Interestingly, participants reflected upon the capability of a human-in-the-loop to mitigate various AI system risks simultaneously. The continuous monitoring of model outputs afforded by human oversight acts as a comprehensive line of defence to identify and prevent system risks propagating into organisational ones. Questions exist over the manner in which human oversight can be best integrated to optimise the inherent capabilities of AI systems and their human supervisors (Cohen et al., 2023). Yet, whilst human oversight can provide a crucial line of defence, the fallibility of human judgement still leaves vulnerability in such systems, necessitating improved organisational learning evidenced in the findings. This level of fallibility rises dramatically in the instance of opaque models (Zetsche et al., 2020). Whilst the crucial importance human oversight other sectors has been argued (De Arteaga, Fogliato and Chouldeck, 2020; Stuurman and Lachaud, 2022; Bakken, 2023), the proliferation of its use discovered in this study asserts its relative importance in a financial context.

### 6.3.2 Testing

The development of robust testing regimes was also identified as a key process within AI RM. Testing has always been an integral element of producing robust software (Tuteja and Dubey, 2012), and thus it is no surprise that this still holds relevance in the context of AI. However, as Poth et al. (2020) note, the field of AI testing is still in development, to which this study adds some nuance. An insightful contribution of this study is illuminating the value of cross-checking models across multiple programming languages and softwares. Due to the opacity of AI systems, participants reflected how models were rigorously tested, aiming for predictability given model inputs; Consistent performance is a critical issue, especially when it comes to higher risk AI systems (Tjoa et al., 2022). The need for continuous testing is also found here, advocated by some authors as a key aspect of AI lifecycle governance (Ortega, Tran and Bandeen, 2023).

### 6.3.3 Structures and Optimisation

Agnese and Capuano (2021) assert the necessity of robust structures for risk governance in FS organisations involving organisational risk teams to facilitate the comprehensive RM at the organisational level. This aligns with the theoretical and widely implemented practical field of enterprise risk management (Bromiley et al., 2015). This study evidences the persistence of these structures in the context of AI, and echoes the hierarchical and structural aspects of existing MRM practices (Hill, 2019; Cosma, Rimo and Torluccio, 2023). The findings portray these structures as decentralised, highlighting the roles of those closest to the models, and organisational risk teams.

The study highlighted how this structure allows comprehensive RM as those closest to the models take the burden of identifying, monitoring and mitigating system risk as they possess the expertise and knowledge of model functioning. Meanwhile, the organisational risk teams guide the deployment and operation of systems, taking into account a holistic picture of organisation wide risk. The study additionally validates the importance of coherent communication to facilitate the cohesive functioning of these structures, whilst adding depth to the practicalities of communicating the mechanics of complex and opaque models. Brown, Steen and Foreman (2009) argue that RM in technologically advanced firms requires a collaborative and communicative approach. The formation of cross-skilled teams seen in these findings is a key aspect of this, and is advocated in other empirical AI RM studies (Madaio et al., 2021; Costanza, Raji and Buolamwini, 2022).

**6.4 Organisational Convergence and Institutional Theory**

The observed convergence of particular RM processes and structures between organisations can, in part, be attributed to institutional theory. This study finds three areas of homogenisation of organisations RM practices: human oversight regimes, testing mandates and decentralised risk governance structures. Participants highlighted how the former two practices conformed to the prevailing regulatory pressures within the finance industry. It is generally agreed that regulatory pressure plays a key part in the development of governance systems (Filatotchev, Jackson and Nakajima, 2013; Koide, 2022), causing organisational convergence through coercive isomorphism.

Model risk regulations are well developed and mandate the need for robust testing and clearly defined roles related to risk within MRM structures (Cosma, Rimo and Torluccio, 2023). The convergence of firms toward particular model oversight structures fosters accountability and aligns with regulatory pressures. In the aftermath of the 2008 global financial crisis, greater pressure was put on accountability, with various jurisdictions including the UK adopting varying degrees of mandated responsibility regimes (Zetsche et al., 2020). Participants noted the capability of human oversight to meet the stringent accountability mandates of the financial regulatory system. These regulatory pressures then shape the development of control systems, leaving human oversight integral to AI RM, and explaining its widespread utilisation. Despite the convergence of the organisations toward these processes and structure, the granular specifics of their implementation was contingent within each organisation.

Despite the apparent applicability of institutional theory in this context. It is capable of explaining only part of the picture of organisational convergence toward particular RM practices. Critically, institutional theory portrays an image of organisational passivity, and is fundamentally at odds with the diverse expectations of organisations underscored by contingency theory. In reality, there are likely a myriad of other internal and external factors underlying firm convergence toward particular practices. Although a number of these were uncovered during data collection, there was not enough evidence to provide their validity across the range of participants surveyed. Thus, although regulatory pressures seemingly play a significant role in organisational convergence, they are undoubtedly not the singular determinant of it.

### 6.5 Practical Implications: Signs of Best Practice

The convergence of these aspects between all the organisations surveyed suggests that these may constitute areas of best practice for AI RM within a financial context (RQ4). Critically, contingency theory undermines the assumption that a single best practice can be universally applied. Thus, drawn from the findings, these recommendations are intentionally foundational and intended to be tailored to context. Furthermore, due to the limited sample, these are tentative guidance as opposed to a rigidly prescribed framework. Figure 3 and Figure 4 display these recommendations from an organisational and process level of risk management respectively.

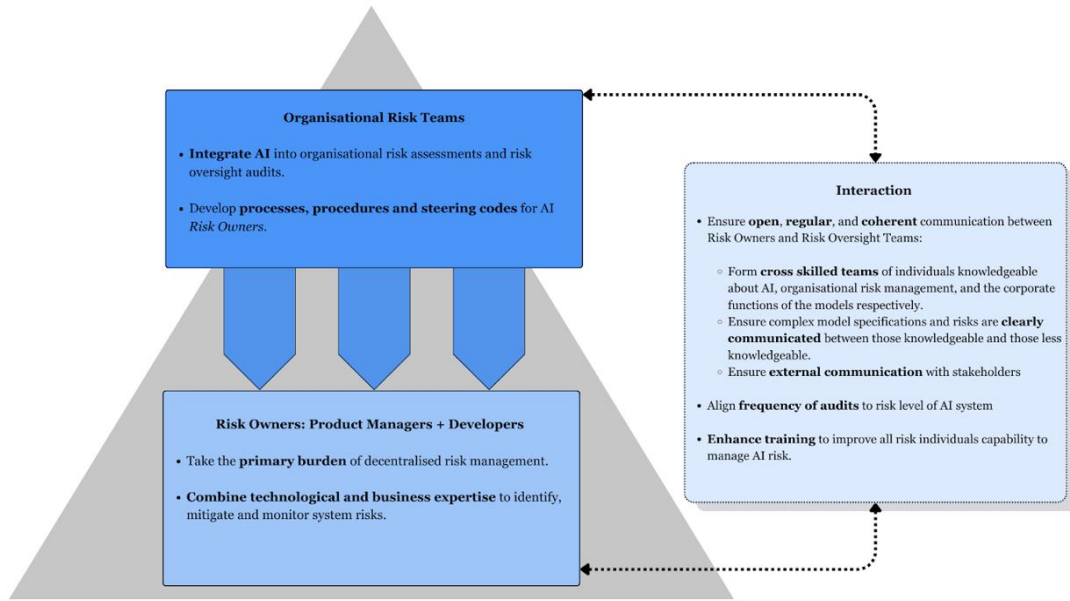

*Figure 3: Risk Management at the Organisational Level*

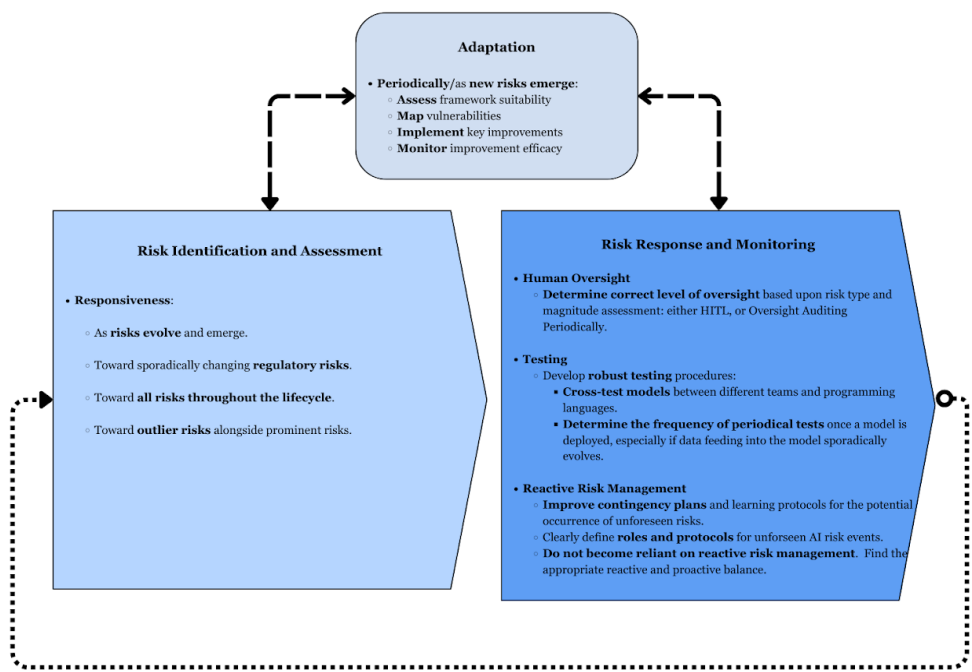

*Figure 4: Risk Management at the Process Level*

## 6.6 Limitations and Research Remedies

In addition to the methodological limitations discussed in Chapter 4, the author would like to note three main holistic limitations of this study, directly aligning them with future research opportunities.

The greatest limitation of this study is potential for the small sample to undermine the credibility of these conclusions through creating internal and external validity issues. Internal and external validity are seen as some of the key qualities of robust qualitative work (Malterud, 2001). Due to the vast implementation potential of AI, a small sample is likely to have left a narrower picture of AI RM. However, as discussed, triangulation techniques or a survey based quantitative or mixed methods methodology were beyond both the scope and constraints of this research. Moreover, a crucial sampling issue was the lack of any organisations implementing high risk AI applications.

However ultimately, the main aim of this study was to provide in-depth and exploratory academic insight into AI RM within FS organisations and not a broad, surface level investigation of the research questions. Thus despite these validity issues, through highlighting emerging patterns of convergence and divergence it has succeeded in providing a deeper understanding of an incipient field. Going forward, the insights here could provide a foundation for a quantitative survey based approach for researchers without the time, knowledge and participant access constraints of the author. This could enable a cross-industry or intra-industry examination of AI RM preparedness and approaches to enable a comprehensive depiction of their current state.

A second key limitation arises due to the scope of the research questions and their ability to provide an encompassing perspective AI RM within the surveyed firms. RQ2 and RQ3 intended to solely find the primary activities and mechanisms employed by organisations, the findings do not depict the entire reality of AI RM and governance for these organisations. Again, despite this endeavour being beyond the scope of this study, the grounding here could enable the research questions to be interrogated in isolation within more extensive studies. To achieve this, the author recommends a case study approach to uncover a more granular picture of AI RM.

A final limitation emerges from the theoretical framing of the study. Guiding the data collection and analysis, the Organisational Risk Management Framework was a novel creation. Embodying the author's preconceptional understanding of organisational RM, its employment may have misaligned the study, omitting other crucial elements of organisational RM. In order to mitigate this, the framework was intentionally left broad whilst being robustly generated from key themes within literature and seminal works, yet future academic interrogation is still needed to assess its validity.

## 6.7 Key Research Avenues

Alongside the aforementioned avenues, this exploratory study has unearthed key areas which require greater scholarly attention.

With respect to vulnerabilities and framework adaptation, a more granular understanding of the areas in which FS organisations' RMF's are susceptible, and the methods and mechanics of RMF adaptation would be beneficial. Classifications of each of these would provide practitioners with guidance to implement a robust AI RMF. A particular issue uncovered here is the importance of reactive strategies. Research into the construction of these strategies would be valuable in the face of escalating emergent AI risks.

As AI RM is in its infancy, many of its constituent activities and mechanisms are still in the process of being fine-tuned. At the process level, scholars could improve methods for assessing risks or investigate optimal constructions of human oversight and testing regimes.

At the organisational level, greater research is needed to understand the optimisation of the decentralised structures through which AI RM is taking place, especially given existing MRM criticism. Meanwhile, existing attempts by academics to incorporate MRM and AI risk into ERM must be advanced to enable a comprehensive and holistic management of AI risks as an overall organisational concern (Scott, Stiles and Debata, 2022).

Due to their potential ramifications, understanding the RM of higher risk AI applications in finance is a particularly pressing issue. Since this study did not come across these use cases, future studies should target these specifically in the effort to refine AI RM. Ultimately, to complement the growing theoretical and practitioner literature, a greater level of empirical research into AI RM is needed across the organisational realm to facilitate a more proficient level of AI RM in practice.

## 7. Conclusion

Adopting a management perspective, the aim of this study was to gain exploratory empirical insight into emerging AI RM practices by investigating an area with relatively mature AI applications, the FS. The findings relay preparedness of RMF's within FS organisations yet uncover the need for targeted adaptation with respect to evolving vulnerabilities. They further identified key approaches and RM elements at both the process and organisational level, which are subsequently formed into a set of tentative best practice recommendations. Alongside this, primary novelty was offered by uncovering the widespread utility of human involvement and the mechanics of distributed structures employed to manage AI risks. At the same time, various themes within the organisational RM and the RM of emerging technologies, software and IT systems were shown to remain relevant in the context of AI.

Overall, in spite of the study's limitations, it offers a valuable perspective into a relatively nascent field.

The implications of this study are twofold. Firstly, it contributes to understanding the RM of novel technologies and information systems. Through unravelling the realities of AI RM in organisations, it complements the extant conceptual literature, and augments the limited empirical literature, on AI RM. Simultaneously, this study offers practical contribution by presenting operationally driven insights for AI RM practitioners. Whilst these findings are nuanced toward the FS, the generalised nature of AI and consistencies between AI applications across industries, provides foundational applicability to a range of contexts. Ultimately, organisational AI RM scholarship is in urgent need to keep pace with the speed of adoption, particularly within the lines of enquiry discussed. Through a coherent effort of future research into the micro level organisational risks, both organisations and wider society can be better safeguarded against the escalating risks of AI.